\documentclass[aps,prb,twocolumn,superscriptaddress,floatfix]{revtex4}
\usepackage{graphicx}
\usepackage{psfrag}
\usepackage{color}
\usepackage{subfigure}
\usepackage{amsmath}
\definecolor{dred}{rgb}{0.7,0.0,0.0}
\allowdisplaybreaks 
 
\begin{document}

\title{Magnetic States of the Two-Leg Ladder \\
Alkali Metal Iron Selenides $A$Fe$_2$Se$_3$ }

\author{Qinlong Luo}
\author{Andrew Nicholson}
\author{Juli\'an Rinc\'on}
\author{Shuhua Liang}
\affiliation{Department of Physics and Astronomy, The University of
  Tennessee, Knoxville, TN 37996, USA} 
\affiliation{Materials Science and Technology Division, Oak Ridge
National Laboratory, Oak Ridge, TN 37831-6138, USA} 

\author{Jos\'e Riera}
\affiliation{Instituto de F\'isica Rosario, Universidad Nacional de Rosario, 2000-Rosario, Argentina}

\author{Gonzalo Alvarez}
\affiliation{Computer Science and Mathematics Division and 
Center for Nanophase Materials Sciences,
Oak Ridge National Laboratory, Oak Ridge, Tennessee 37831, USA}

\author{Limin Wang}
\affiliation{Condensed Matter Physics and Materials Science Department,
Brookhaven National Laboratory, Upton, New York 11973, USA}

\author{Wei Ku}
\affiliation{Condensed Matter Physics and Materials Science Department,
Brookhaven National Laboratory, Upton, New York 11973, USA}
\affiliation{Physics Department, State University of New York, Stony Brook,
New York 11790, USA}

\author{German D. Samolyuk}

\affiliation{Materials Science and Technology Division, Oak Ridge 
National Laboratory, Oak Ridge, TN 37831-6138, USA}

\author{Adriana Moreo}
\author{Elbio Dagotto}

\affiliation{Department of Physics and Astronomy, The University of
  Tennessee, Knoxville, TN 37996, USA} 
\affiliation{Materials Science and Technology Division, Oak Ridge
National Laboratory, Oak Ridge, TN 37831-6138, USA}

\date{\today}

\begin{abstract}
Recent neutron scattering experiments 
addressing the magnetic state of the two-leg ladder
selenide compound BaFe$_2$Se$_3$  have unveiled a dominant spin arrangement 
involving ferromagnetically ordered
2$\times$2  iron-superblocks, that are antiferromagnetically coupled 
among them (the ``block-AFM'' state). Using the 
electronic five-orbital Hubbard model, first principles techniques
to calculate the electronic hopping amplitudes between irons, and
the real-space Hartree-Fock approximation to handle the many-body effects, 
here it is shown that the exotic block-AFM state is
indeed stable at realistic electronic densities close to $n \sim 6.0$. 
Another state (the ``CX'' state) with parallel spins along the rungs
and antiparallel along the legs of the ladders is close in energy.
This state
becomes stable in other portions of the phase diagrams, such as with hole
doping,
as also found experimentally via neutron scattering 
applied to KFe$_2$Se$_3$. In addition, 
the present study unveils other competing magnetic phases 
that could be experimentally stabilized varying either $n$ chemically or the 
electronic bandwidth by pressure. Similar results
were obtained using two-orbital models, studied here via Lanczos 
and DMRG techniques. A comparison of the results obtained with the
realistic selenides hoppings amplitudes for BaFe$_2$Se$_3$ against 
those found using the hopping
amplitudes for pnictides reveals several qualitative similarities,
particularly at intermediate and large Hubbard couplings.


\end{abstract}

\maketitle

\section{Introduction}

The study of Fe-based superconductors~\cite{johnston,recent-review,recent-rmp} 
continues unveiling fascinating new discoveries at a fast pace.
Among the most recent developments 
is the report of superconductivity in the intercalated iron
selenides  K$_{0.8}$Fe$_{2-x}$Se$_2$ 
and (Tl,K)Fe$_{2-x}$Se$_2$.\cite{guo} In addition, 
at the special composition K$_{0.8}$Fe$_{1.6}$Se$_2$ with
the iron vacancies in a $\sqrt{5}\times\sqrt{5}$ arrangement, neutron scattering
studies\cite{bao,fe} of this (insulating) compound
have revealed an unusual magnetic order. This magnetic state
involves 2$\times$2 iron blocks with their four spins
ferromagnetically ordered, large ordering temperatures, and 
concomitant large magnetic moments $\sim$3.3~$\mu_B$/Fe. 
The 2$\times$2 blocks are
antiferromagnetically coupled among them. Phase
separation tendencies have also been reported 
in this type of insulators.\cite{ricci}
Photoemission experiments for (Tl, K)Fe$_{1.78}$Se$_2$
revealed a Fermi surface  with only
electron-like pockets at wavevectors $(\pi,0)$ and $(0,\pi$),\cite{ding} 
showing that the Fermi surface nesting of hole and electron
pockets is not sufficient to understand these materials.\cite{recent-review}

The developments described above suggest that progress in
the understanding of chalcogenides could
be made if the iron spins are arranged differently than in the nearly square 
lattice geometry
of the  FeSe layers. For this reason considerable interest was generated
by recent studies\cite{first,caron,petrovic,athena,caron2,nambu} 
of BaFe$_2$Se$_3$ (the ``123'' compound) since this material contains chains made of
[Fe$_2$Se$_3$]$^{2-}$ building blocks separated by Ba.
These effective two-leg iron ladders  
in BaFe$_2$Se$_3$ are cut-outs 
of the layers of edge-sharing FeSe$_4$ 
tetrahedra normally found in layered chalcogenides. 
Each double chain consists of pairs of
 iron atoms (the ``rungs'') located one next to the other forming 
a one dimensional arrangement perpendicular to those rungs, 
defining indeed a two-leg
 ladder structure.
In the context of the Cu-oxide high-T$_c$ superconductors, spin 1/2 two-leg 
ladders have also been much studied because of their unusual spin gap, induced  
by the ladder geometry. The Cu-oxide-ladder spin state 
is dominated by rung spin-singlets, 
and a tendency to superconduct upon doping.\cite{ladders-original,dagotto-rice} 
In particular, the compound
SrCu$_2$O$_3$ is the Cu-based analog of BaFe$_2$Se$_3$.\cite{dagotto-ladder}

A recent remarkable development that increases the relevance of the
iron-based ladders is the following. The preparation of a single layer of
alkali-doped FeSe with the geometry of weakly coupled
two-leg ladders was recently reported in Ref.~\onlinecite{SCladder},
where it was also argued that this ladder system is 
superconducting based on the presence
of a gap in the local density of states. These results suggest that Fe-based 
ladders provide
a simple playground where even superconductivity can be explored, 
increasing the similarities with the Cu-oxide ladders that are
also superconducting.\cite{dagotto-ladder}

\begin{figure}[thbp]
\begin{center}
\includegraphics[width=7.0cm,clip,angle=0]{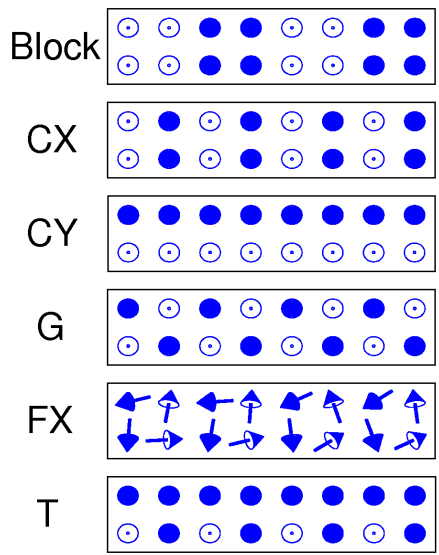}
\vskip -0.3cm
\caption{(Color online) 
Magnetic states observed in the phase diagrams of the
multiorbital Hubbard models used in this study, employing the geometry
of a two-leg ladder. 
}
\vskip -0.5cm
\label{fig1}
\end{center}
\end{figure}

BaFe$_2$Se$_3$  is an insulator, with a resistivity 
displaying an activation 
energy between $\Delta$$\sim$0.13~eV (Ref.~\onlinecite{nambu}) and
$\Delta$$\sim$0.178~eV (Ref.~\onlinecite{petrovic}). 
The 123-ladder compound has long-range antiferromagnetic (AFM) 
order at $\sim$250~K, with low-temperature 
magnetic moments $\sim$2.8~$\mu_B$, and it displays short-range
AFM correlations at higher temperatures (in particular 
$\xi$$\sim$35$\rm \AA$ at room-$T$).\cite{caron,petrovic,athena} 
Upon cooling, the magnetic order presumably settles along the
ladder directions first, and then weaker interladder interactions establish the 
long-range order. Neutron diffraction studies\cite{caron,nambu} 
reported a dominant magnetic 
order at low-$T$ involving blocks
of four iron atoms with their moments aligned, coupled antiferromagnetically 
along the ladder
direction (Fig.~1, top state). This state is 
sometimes dubbed the plaquette state, but
here it will be referred to as the ``block-AFM'' state or just ``Block''. 
The ferromagnetic 2$\times$2 
building blocks present in the block-AFM state of the ladders
are the same blocks reported before in K$_{0.8}$Fe$_{1.6}$Se$_2$, with
the iron vacancies in the $\sqrt{5}$$\times$$\sqrt{5}$ distribution.
When the 123-ladder material is doped with K as in Ba$_{1-x}$K$_x$Fe$_2$Se$_3$, 
experimentally it is known that the magnetic state evolves from the block-AFM state,
through a spin glass, eventually arriving for KFe$_2$Se$_3$
to the spin state labeled ``CX'' also displayed in Fig.~\ref{fig1},
where the spins in the same rung are coupled ferromagnetically but they are 
antiferromagnetically ordered in the long ladder direction.\cite{caron2} 
Note that in BaFe$_2$Se$_3$ the valence of Fe is expected to be 2+, if
those of Ba and Se are +2 and -2, respectively, giving an electronic density $n$=6.0.
But in KFe$_2$Se$_3$, K has valence +1, thus rendering the average valence of
Fe to be +2.5, that corresponds to an electronic density $n$=5.5.

In the present publication, results for multiorbital Hubbard models 
are reported. The lattice distortions\cite{caron,petrovic} 
are partially taken into account
via the hopping amplitudes, as described below.
However, part of our results presented in the following sections show that 
even without lattice distortions 
the 2$\times$2 block-AFM state is stable 
in regions of the phase diagrams that are constructed by varying
the on-site Hubbard repulsion $U$, the Hund coupling $J_{\rm H}$, 
and the electronic density $n$.\cite{compatible} In other words,
our most important result is that several models, studied with several
approximations, systematically contain the block-AFM state
as a robust phase in the phase diagram.
Moreover, the other
recently observed\cite{caron2} CX-state is also 
found in the resulting phase diagrams. Several
competing states that could be stabilized in 
related compounds or under pressure or via chemical doping are also discussed. 

The present study is carried out mainly using the Hartree-Fock 
approximation for the five-orbital Hubbard model, 
employing both a set of hopping amplitudes that are deduced
from first principles techniques applied to the selenide 123 ladders, 
as explained below, and also an ``old''
set of hopping that was previously employed in layered pnictides. The
purpose of using two sets of hoppings is 
to gauge how sensitive the results are 
with regards to modifications in those hopping amplitudes. In addition,
our results for pnictide hoppings can be considered predictions in
case two-leg ladder pnictides are synthesized in the future. 
Results for a reduced two-orbital Hubbard model using Lanczos\cite{lanczos} and the
Density Matrix Renormalization Group (DMRG)\cite{dmrg} techniques 
are also presented here, also for two sets
of hoppings. Overall, the two phases observed experimentally in
neutron scattering, the block-AFM and the CX states, are stable in regions of
the phase diagram centered at the realistic Hund coupling $J_{\rm H}/U=0.25$. 
With regards to phase diagrams, gaps, magnetic moments, and competing states, 
a reasonable qualitative agreement is found between 
the two different hopping sets, and for the
different number of orbitals considered in our effort.
Studies of the spin-fermion two-orbital model\cite{spin-fermion1} using 
Monte Carlo simulations\cite{spin-fermion2} also 
provide a phase diagram compatible with
those of the Hubbard models.

\section{Five-orbital Hubbard model, hopping amplitudes, and methods} 

In this section, the focus will be on the derivation of the hopping amplitudes
needed for the five-orbital Hubbard model and in providing details of the
real-space Hartree-Fock technique employed.
The models used in this publication have all
been extensively discussed before, thus details will not be repeated. In particular,
the five-orbital Hubbard model is explicitly defined in Ref.~\onlinecite{luo}. 
With regards to the hopping
amplitudes for the 123-ladder compounds, here they have been calculated using first-principles
techniques, as explained below. These hoppings will be refereed 
to as the selenide hoppings
in the rest of the publication.\cite{imada}
For completeness, results using the
hoppings corresponding to layered pnictides~\cite{graser}
will also be used, and the results compared with one another. 
While the data gathered with the realistic selenide hoppings 
are our most important set of results, contrary to naive expectations it will
be shown that a reasonable agreement is observed between these two ``a priori''
quite different sets of hopping amplitudes,  at least at a qualitative level. 
The electronic density of main interest 
is, in principle, $n\sim6.0$ (i.e. 6 electrons/Fe), thus our efforts 
are centered at this density, 
but some results varying $n$ are shown below as well (or verbally described).
As explained before, 
the on-site intraorbital Hubbard repulsion is $U$, the Hund coupling is $J_{\rm H}$,
and the interorbital repulsion $U'$ is assumed to satisfy $U'$=$U$-2$J_{\rm H}$. Ladders 
of sizes 2$\times$$L$ ($L$=4, 8, 16, 32) were studied, and size effects were found to be mild.
Periodic (open) boundary conditions are used along the chain (rung) direction.

\begin{figure}[thbp]
\begin{center}
\includegraphics[width=7.0cm,clip,angle=0]{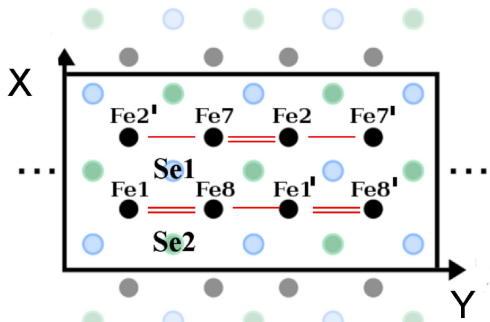}
\vskip -0.3cm
\caption{(Color online) 
Label convention of the iron sites used in Table I, adapted from Fig.~1(b)
of Ref.~\onlinecite{caron}. The single and double
lines along the $y$ axis denote two different lattice spacings, with
specific numbers taken from Ref.~\onlinecite{caron}. The two selenium 
sites denote locations above and below the plane defined by the iron
ladder.
}
\vskip -0.5cm
\label{fig1.bis}
\end{center}
\end{figure}

\begin{table}
\begin{ruledtabular}
\begin{tabular}{|c|c|}
Matrix & BaFe$_2$Se$_3$\\ \hline
$t^{\rm On Site}$ &
$\begin{pmatrix}
-0.4604 & -0.0617 & ~0.0534 & -0.0345 & -0.0178\\
-0.0617 & -0.5947 & -0.0851 & ~0.0371 & ~0.0169\\
~0.0534 & -0.0851 & -0.0719 & -0.0030 & ~0.0165\\
-0.0345 & ~0.0371 & -0.0030 & -0.1669 & ~0.0286\\
-0.0178 & ~0.0169 & ~0.0165 & ~0.0286 & -0.1632
\end{pmatrix}$ \\ \hline
$t^{\rm NN}_{{\rm leg}, {1 \rightarrow 8}}$&
$\begin{pmatrix}
-0.0807   & -0.3276   & -0.0139   & ~0.2734   & ~0.0456  \\
-0.3276   & -0.2875   & -0.0702   & ~0.2661   & ~0.0228  \\
~0.0139   & ~0.0702   & -0.1477   & -0.0531   & ~0.2714  \\
~0.2734   & ~0.2661   & ~0.0531   & -0.2733   & ~0.0373  \\
-0.0456   & -0.0228   & ~0.2714   & -0.0373   & -0.0397
\end{pmatrix}$ \\ \hline
$t^{\rm NN}_{{\rm leg}, {1' \rightarrow 8}}$&
$\begin{pmatrix}
~0.0497   & -0.2674   & ~0.0187   & -0.1186   & -0.0738  \\
-0.2674   & -0.3943   & -0.0388   & -0.3449   & -0.0195  \\
-0.0187   & ~0.0388   & -0.0580   & ~0.0199   & -0.2689  \\
-0.1190   & -0.3449   & -0.0199   & -0.3107   & -0.0147  \\
~0.0738   & ~0.0195   & -0.2689   & ~0.0147   & -0.1343
\end{pmatrix}$ \\ \hline
$t^{\rm NN}_{{\rm rung}, {8 \rightarrow 7}}$ &
$\begin{pmatrix}
-0.0421 & ~0.2853 & -0.1718 & -0.0162 & ~0.0055\\
~0.2853 & -0.3801 & ~0.3311 & ~0.0411 & ~0.0098\\
-0.1718 & ~0.3311 & -0.2881 & -0.0115 & -0.0259\\
-0.0162 & ~0.0411 & -0.0115 & -0.0058 & -0.2303\\
~0.0055 & ~0.0098 & -0.0259 & -0.2303 & -0.0153
\end{pmatrix}$ \\ \hline
$t^{\rm NN}_{{\rm rung}, {1 \rightarrow 2'}}$ &
$\begin{pmatrix}
-0.0421 & ~0.2853 & ~0.1718 & -0.0162 & -0.0055\\
~0.2853 & -0.3801 & -0.3311 & ~0.0411 & -0.0098\\
~0.1718 & -0.3311 & -0.2881 & ~0.0115 & -0.0259\\
-0.0162 & ~0.0411 & ~0.0115 & -0.0058 & ~0.2303\\
-0.0055 & -0.0098 & -0.0259 & ~0.2303 & -0.0153
\end{pmatrix}$ \\ \hline
$t^{\rm NNN}_{2 \rightarrow 8}$&
$\begin{pmatrix}
-0.0185 & ~0.0054 & ~0.1140 & ~0.0893 & -0.0721\\
-0.0159 & -0.0379 & -0.0661 & ~0.0603 & -0.0118\\
-0.1483 & ~0.0837 & ~0.2117 & ~0.0843 & ~0.0679\\
-0.1442 & -0.0490 & ~0.0801 & ~0.1823 & ~0.0559\\
-0.0879 & -0.0017 & -0.0304 & -0.0529 & ~0.0644
\end{pmatrix}$ \\ \hline
$t^{\rm NNN}_{2' \rightarrow 8}$&
$\begin{pmatrix}
-0.0185 & ~0.0054 & -0.1140 & ~0.0893 & ~0.0721\\
-0.0159 & -0.0379 & ~0.0661 & ~0.0603 & ~0.0118\\
~0.1483 & -0.0837 & ~0.2117 & -0.0843 & ~0.0679\\
-0.1442 & -0.0490 & -0.0801 & ~0.1823 & -0.0559\\
~0.0879 & ~0.0017 & -0.0304 & ~0.0529 & ~0.0644
\end{pmatrix}$ \\ \hline
\end{tabular}
\end{ruledtabular}
\vskip 0.2cm
\caption{\label{tablehops5} Hopping matrices for the
BaFe$_2$Se$_3$ material obtained from a tight-binding Wannier function analysis of
the first principles results (in eV units). 
The matrices are written in the orbital basis
$\{d_{z^2},\, d_{x^2-y^2},\, d_{yz},\, d_{xz},\, d_{xy}\}$ for the
on-site energy and inter-orbital hopping ($t^{\rm On Site}$), 
nearest-neighbors hoppings ($t^{\rm
NN}$, both along the rungs and the legs) and next-nearest-neighbors hoppings  ($t^{\rm NNN}$). 
The long (short) direction of the ladder 
is oriented along the $y$ ($x$) axis. The convention for the 
iron site labels is in Fig.~\ref{fig1.bis}. Note that each 5$\times$5 matrix in this table should be
considered as the hopping matrix to move from one iron to another as indicated. For a given
Fe-Fe bond, the full matrix
that includes both the back and forth processes for the hopping is of size 10$\times$10 and it
consist of a 5$\times$5 matrix of 
this table in a non-diagonal block, the transpose in the other 
nondiagonal 5$\times$5 block, and the on-site matrix (top of this table) in both diagonal blocks.}
\end{table}

The selenides hopping amplitudes for the 123 ladders
were obtained via a first-principles density functional theory calculation of the non-magnetic
normal state. The calculation was conducted using the WIEN2K  
implementation of the full potential linearized augmented plane wave method in the
local density approximation.\cite{blaha} 
 The $k$-point mesh was taken to be 7$\times$15$\times$19. The lattice constants were
taken from Ref.~\onlinecite{caron}.
Table I shows the hopping parameters calculated by representing the resulting self-consistent Kohn-Sham
Hamiltonian with low-energy ([-2.5, 2] eV) symmetry-respecting Wannier functions~\cite{wannier}
with strong Fe-$d$ symmetry.
Since the influence of the As-$p$ orbitals are integrated into the tail of the Wannier functions, the parameters
correspond to an effective iron-only model with 
five orbitals per iron. The 
staggered location of the selenium atoms, above and below the plane defined
by the Fe atoms, is taken into account in the calculation. 
For a similar discussion in the context
of the three orbital model, see Ref.~\onlinecite{3orbitals}.
Note also that Table I contains
the hoppings that are needed for the full description of the system, based
on the iron locations in Fig.~\ref{fig1.bis}. Other hoppings are all identical
to one of those shown in Table I. For instance, the hopping matrix from Fe2 to Fe7
is the same as the hopping matrix from Fe1 to Fe8 in Table I, the hopping matrix
from Fe1 to Fe7 is the same as the hopping matrix from Fe2 to Fe8, etc.


In addition, and for completeness, the hopping parameters of Ref.~\onlinecite{graser} obtained for pnictide compounds
were also used in our studies. The goal was to test our conclusions against reasonable modifications in the hoppings.
Several aspects of 
our results were found to be 
qualitatively similar for the two sets of hoppings, although
of course, quantitatively there are substantial differences.

The five-orbital Hubbard model is studied here using the
real-space Hartree-Fock (HF) approximation, taking into account the
staggered location of the Se atoms via the proper hopping amplitudes. Recently, the same method 
was successfully employed in the analysis of K$_{0.8}$Fe$_{1.6}$Se$_2$
and other systems.\cite{luo-245,luo-stripes} 
A real-space approach, where all the HF expectation values that need to be
found self-consistently are assumed independent from site to site, allows
for the system to select spontaneously the state that minimizes the HF energy,
reducing the bias into the calculations. The HF expectation
values are obtained by an iterative process that reduces the energy until
convergence.\cite{numerical} The results shown below
have been computationally obtained by two procedures: (i) using as initial 
configurations the states in Fig.~\ref{fig1} and then comparing energies after convergence, 
or (ii) starting with totally random values and analyzing the spin order
particularly at short distances. Procedure (ii), computationally demanding due to the
large number of iterations needed, sometimes allows for the identification
of ``unexpected'' phases [that then become part of (i)] and also to confirm the
results of procedure (i).

\section{Results for the five-orbital Hubbard model}

\subsection{Phase diagrams and the block-AFM phase}

\begin{figure}[thbp]
\begin{center}
\includegraphics[width=8.8cm,clip,angle=0]{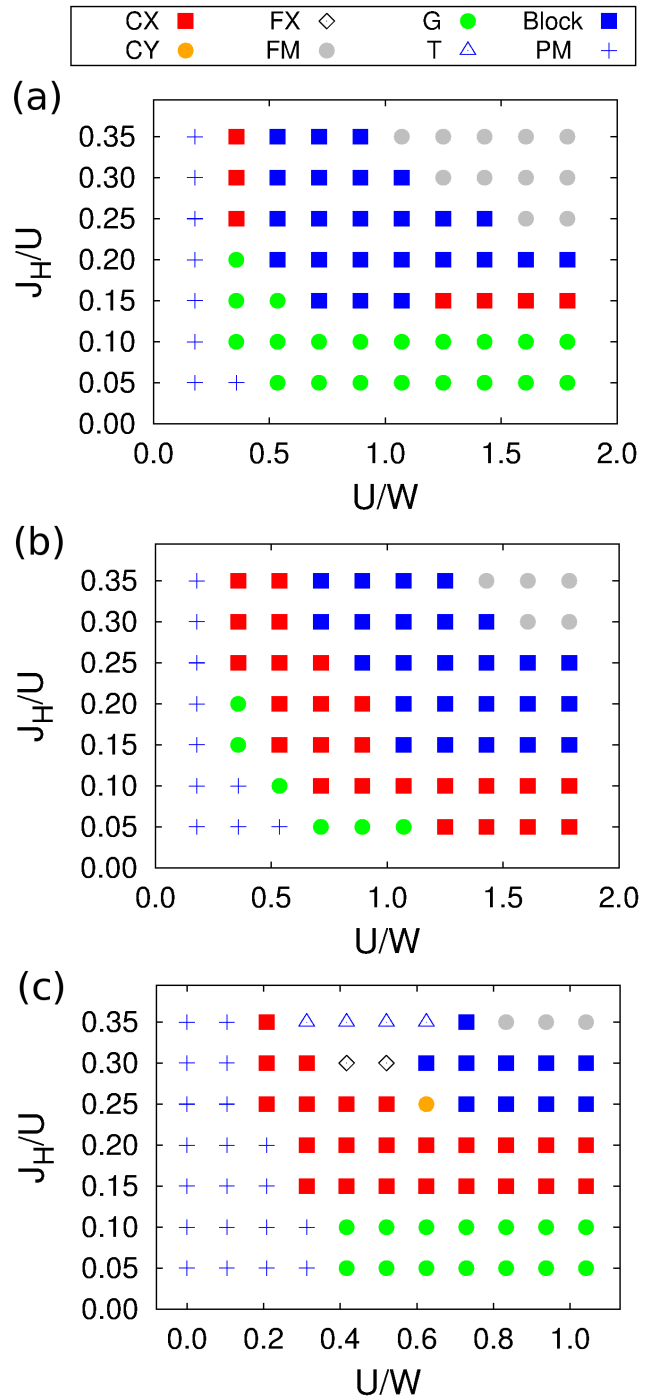}
\vskip -0.3cm
\caption{(Color online) 
Phase diagram of the five-orbital Hubbard model in the real-space HF approximation.
The label convention for the phases is in the upper inset and also in Fig.~\ref{fig1}. 
PM denotes a paramagnetic state.
(a) Results for a 2$\times$16 cluster, using 
the selenides hopping amplitudes for the 123 ladders, and at 
electronic density $n=6.0$.
(b) Same as (a) but for electronic density $n=5.75$. 
(c) Results for a 2$\times$32 cluster, using 
the pnictides hopping amplitudes, and working at electronic density $n=6.0$.
}
\vskip -0.4cm
\label{fig2}
\end{center}
\end{figure}

The main results of the present HF study of the five-orbital Hubbard model
are shown in Fig.~\ref{fig2}, where the phase diagrams varying $U/W$ 
and $J_{\rm H}/U$ are presented using both the realistic 
selenides hoppings for the 123 ladders as well as the pnictides hoppings for comparison.
The bandwidth $W$ of the five-orbital Hubbard model is $\sim$2.8~eV for the
selenides hoppings, while for the
pnictides hoppings it is $\sim$4.8~eV.\cite{graser}
In all cases, Figs.~\ref{fig2}(a-c) show the remarkable result
that the block-AFM phase found in neutron experiments
for the 123 ladders\cite{caron,nambu} becomes stable in a robust region
of the phase diagram. This is interesting since the 2$\times$2 
blocks in ladders are not as clearly geometrically defined as  
in the $\sqrt{5}\times\sqrt{5}$ iron-vacancies 
arrangements, where each of the plaquettes 
of the tilted square lattice of those iron vacancies already 
contains a 2$\times$2 block inside. In our two-leg ladders, on the other hand, 
the ferromagnetic (FM) blocks do emerge spontaneously 
in the calculations described here and in experiments as well. 
As explained before, the 
present results were confirmed using 
unbiased random starting configurations for the HF expectation
values and an iterative procedure for convergence. By this procedure, 
the stability of the block-AFM state was indeed tested
at several points of the phase diagram.
Moreover, it is interesting that the region of stability 
includes the realistic ratio $J_{\rm H}/U=0.25$, found before
to correspond to the ``physical region'' where a good agreement 
theory-experiment was observed for the pnictides.\cite{luo,luo-extra}

With regards to the actual value of $U/W$, note that the block-AFM phase is
stabilized starting at $U/W \sim 0.5-0.6$ for the selenides
hoppings [Figs.~\ref{fig2}(a,b)] and at $U/W \sim 0.6$ for 
the pnictide hoppings [Fig.~\ref{fig2}(c)]. 
This is similar to the value $\sim$0.52 reported
for K$_{0.8}$Fe$_{1.6}$Se$_2$ in Ref.~\onlinecite{luo-245} using similar
techniques. The critical $U/W$'s quoted above are slightly larger than 
the $U/W \sim 0.31$ needed for the pnictides 
in the planar geometry of the ``1111'' and ``122'' materials to form
the C-type AFM state,\cite{luo,luo-extra} but note that in our present results
magnetic order in the CX channel (the analog of the C-type AFM phase) is reached at
$U/W \sim 0.3$ in good agreement with those previous investigations.
Considering that it is the block-AFM state that is found experimentally 
for the 123-ladder 
selenides, this suggest that these selenides are more strongly correlated than pnictides
and their ratios of $U/W$ are roughly 0.5/0.3 = 1.66.
Note also that the actual values of $U/W$ are still smaller than 1, 
the ratio often considered as the boundary
of the strong coupling limit, implying that the selenide ladders
are still ``intermediate'' coupling compounds.\cite{recent-review} 
However, the HF approximation favors ordered states, 
and including quantum fluctuations the $U/W$
needed to stabilize the block-AFM phase may exceed 1. 
On the other hand, note also that results in real ladders may be influenced by the
presence of a robust electron-lattice coupling (mentioned in Refs.~\onlinecite{caron,petrovic}) 
that may render stable the block-AFM phase even
at values of $U/W$ not as large as needed for its stabilization when based entirely on 
an electronic mechanism. In spite of these caveats, 
it is clear that even with lattice
distortions incorporated the presence of sizable electronic correlations
appears to be important to stabilize the block-AFM state.

In the block-AFM phase found in our study the magnetic moment per Fe is
large and close to saturation. More specifically, it is 
$\sim$4.0~$\mu_B$ for the selenides hoppings and
$\sim$3.9~$\mu_B$ for the pnictides hoppings, with small variations caused
by the selection of specific values of $U$.
The difference with the experimental result\cite{caron} $\sim$2.8~$\mu_B$ 
may be caused by the absence of fluctuations 
in the HF approximation,\cite{fluctuations-cite} or by the neglect of lattice
distortions in our effort, as already discussed. But at least qualitatively the
large value of the magnetic moment, as compared with the relatively small moments
reported in some layered pnictides, is here properly reproduced.

In the present Hartree Fock effort, square 8$\times$8 clusters 
have also been studied to address the
coupling between ladders in the direction perpendicular to the legs.
In practice, a weak interladder coupling was introduced 
by  multiplying by a small factor $\alpha$=0.1 
all the hoppings connecting sites belonging
to different individual 2$\times$8 ladders (thus the 8$\times$8 cluster has four
of these two-leg ladders). Other values of (small) $\alpha$
were used and the results were all similar. The main result (not shown) 
is that the phase diagrams using the 8$\times$8 clusters are virtually 
identical to those found for the individual
two-leg ladders, for both sets of hoppings, with the only 
interesting detail that the weak coupling between the ladders establishes
an effective antiferromagnetic coupling between them, 
as found experimentally.\cite{caron}

\subsection{The CX phase and other competing states}

It is important to remark that 
in all Figs.~\ref{fig2}~(a-c) there are several 
other magnetic states in addition to the block-AFM state. In 
particular, the ``CX'' phase found experimentally 
in hole-doped ladders\cite{caron2} also occupies
a robust region of the phase diagram, and 
it is located next to the block-AFM phase for both the selenides 123-ladder hoppings
as well as the pnictide hoppings, at the electronic densities 
investigated in Fig.~\ref{fig2}. Its region of stability 
includes areas with smaller or
similar values for $U/W$ than those where the block-AFM 
state is stable. Our investigations
varying $n$ reveal that this phase is stable in a broad region of parameter
space, including the $n=5.5$ electronic
density corresponding to KFe$_2$Se$_3$,\cite{caron2} indicating once 
again a good agreement between calculations and experiments. In fact, Fig.~\ref{fig2}~(b)
shows that the CX state is more stable at electronic density $n$=5.75 than at $n$=6.0,
compatible with experiments.
The CX state can be considered closely related to the C-AFM state of layered pnictides
with the wavevector ($\pi$,0), thus its stability particularly close to the PM state 
should not be too surprising.

Varying $U/W$ and $J_{\rm H}/U$, phases that have not been observed experimentally for the
two-leg ladders become stable. 
For instance, when the Hund coupling is small compared with
$U$, a G-type antiferromagnetic state is found, with staggered magnetic order. 
In the other extreme of magnetic order,
ferromagnetism is observed in a small region of parameter space 
for a sufficiently large $U$ and Hund couplings, for both sets
of hoppings. The qualitative tendency from ``G'' to ``CX'' to ``block-AFM'' to ``FM'' with
increasing $J_{\rm H}/U$ at robust $U/W$ goes together 
with the tendency to FM order in the vicinity of each iron atom: 
for the G-state the three NN links are AFM, for the CX-state two are AFM 
and one is FM, for the block-AFM state two are FM and
one AFM, and of course for the full FM state all NN links are FM.

Small ``islands'' of other
states are also present in the phase diagram corresponding to the pnictides hoppings, 
including the CY-state which is another
relative of the C-AFM state of the pnictides, 
as well as the Flux and T states 
(see Fig.~\ref{fig1} for the spin arrangement
corresponding to these states). But it is clear that the block-AFM,
CX, G, FM, and PM states dominate the phase diagrams.

\subsection{Density of States}

\begin{figure}[thbp]
\begin{center}
\includegraphics[width=8.5cm,clip,angle=0]{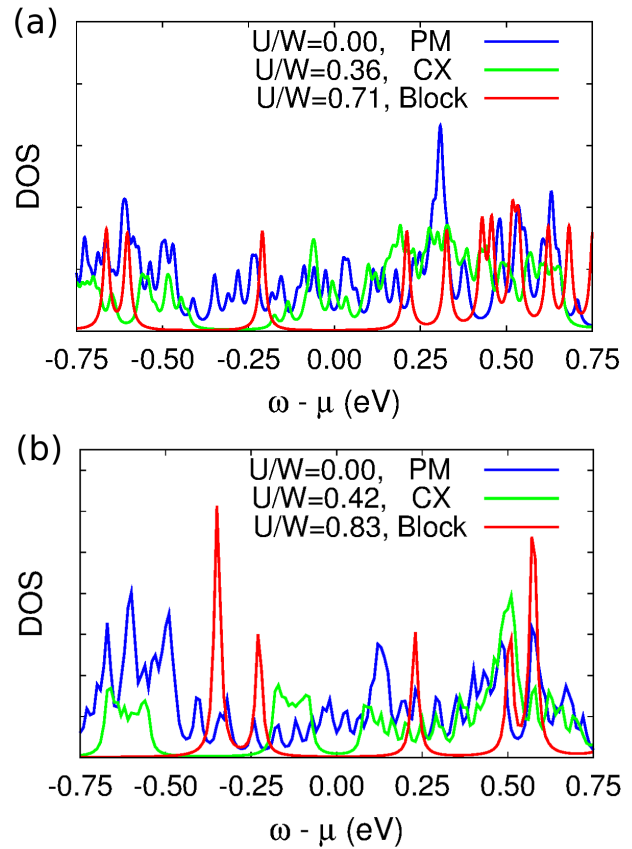}
\vskip -0.3cm
\caption{(Color online) 
Density of states of the five-orbital Hubbard model (in the HF approximation),
at $J_{\rm H}/U=0.25$, and the values of $U/W$ indicated. 
The type of phase state corresponding to each
value of $U/W$ is also indicated. 
(a) corresponds to the selenides hoppings for the 123-ladder 
compound and electronic density $n=6.0$. The bandwidth $W$ in 
this case is $\sim$2.8~eV.
(b) corresponds to the pnictides hoppings, for comparison. 
The electronic density is $n=6.0$,
and the bandwidth $W$ is $\sim$4.8~eV.
In both cases, the small oscillations at $U/W=0$ and in the CX phase 
are caused by size effects in the long direction 
of the 2$\times$16 or 2$\times$32 clusters used and the 
intrinsic small size in the rung direction. 
}
\vskip -0.4cm
\label{fig3}
\end{center}
\end{figure}

The density of states (DOS) 
of the block-AFM state for both the cases of the ``selenides'' 123-ladder 
hoppings and the ``pnictides'' hoppings
are shown in Fig.~\ref{fig3} for representative couplings. 
The presence of a gap
at the chemical potential for the block-AFM state 
and for both hoppings indicates an insulating state, 
in agreement with experiments. While
the values of the gap for the block-AFM state 
($\Delta$$\sim$0.40~eV and $\Delta$$\sim$0.45~eV for
the selenides and pnictides hoppings, respectively) 
are larger than reported experimentally,\cite{nambu,petrovic}
the qualitative trends are correct. Further improvement with
experiments can be achieved by better fine tuning $U/W$ and $J_{\rm H}/U$,
by adding effects arising from the three dimensionality of the problem,
incorporating other lattice distortions, etc.
 

Another detail that merits a comment is that the CX phase, being 
closer in the phase diagrams to the PM state than the block-AFM state is,
has a metallic or weakly insulating character that depends on specific
details such as the value of $U/W$. At a fixed $J_{\rm H}/U$ such
as 0.25, the metal-insulator transition seems to occur within the CX phase.

\section{Results for the two-orbital Hubbard model}

The results obtained via the HF approximation to the five-orbital Hubbard model
can be further analyzed, at least qualitatively, 
by studying models with less orbitals but using computational
techniques beyond the mean-field approximation. 
Consider for instance the two-orbital Hubbard
model employing the $d_{xz}$ and $d_{yz}$ orbitals. 
This model can be studied exactly in small clusters, via the Lanczos
algorithm,\cite{lanczos} or via DMRG techniques. Here, 
in the first part of the study
using the two-orbital model, the 
hoppings originally developed for pnictides will be used, 
but then in the second part a set of hoppings derived from 
the five-orbital hoppings of the ladder selenides will be employed. For
details about the model, particularly the hoppings for the pnictides case, and for the 
technical aspects of the implementation
of the Lanczos technique to multiorbital models
see Refs.~\onlinecite{andrew,andrew1,andrew2}.
A phase diagram can be constructed by
calculating the spin-structure factor $S(q_x,q_y)$, 
and focusing on the wavevector that maximizes
this quantity. In particular, if $S(q_x,q_y)$ is maximized
at (0,0) then the state is considered ferromagnetic, if maximized at ($\pi$,0) it is CX,
if at (0,$\pi$) it is CY, and if at ($\pi$/2,0) it is the block-AFM state. Before
providing the results, it is important to clarify that the discussion involving
two orbitals is {\it qualitative} at best. In fact, the two-orbitals model version
of the five-orbitals model derived here from first principles provides only a limited fit of the
Fermi Surface. However, it will be shown that in spite of this quantitative issues, the
crude two-orbitals model used here do present the block-AFM and CX states in the phase
diagrams, showing that, within reason, their presence does not depend on details 
of the hopping amplitudes.

\begin{figure}[thbp]
\begin{center}
\includegraphics[width=8.5cm,clip,angle=0]{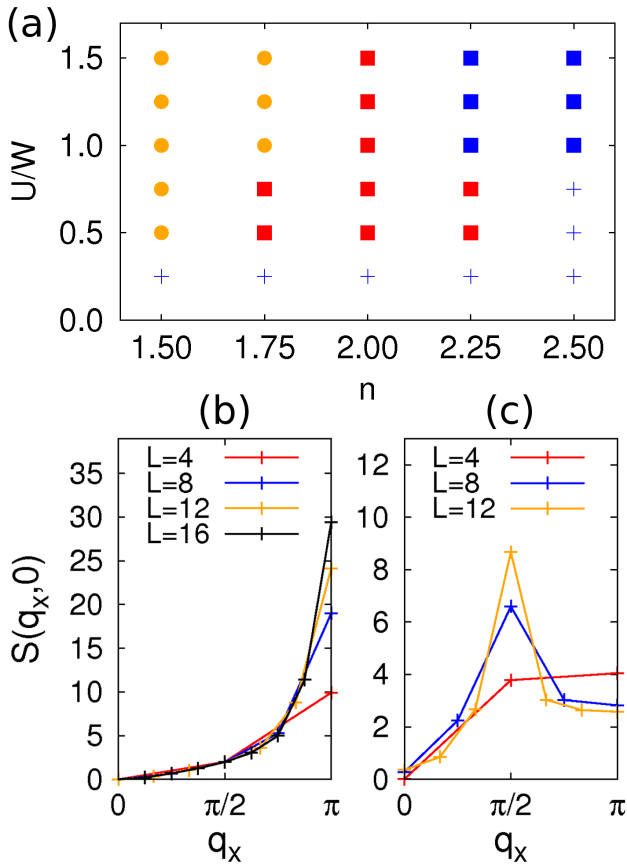}
\vskip -0.3cm
\caption{(Color online) 
(a) Phase diagram of the two-orbital Hubbard model studied
with the Lanczos method on a 2$\times$4 cluster, 
and at $J_{\rm H}/U=0.25$. 
Shown are results varying $U/W$ ($W=12$~eV, see Ref.~\onlinecite{andrew}) 
and the number of total electrons
$N$ as 12, 14, 16 ($n=2$), 18, and 20. The assignments of the many phases
are decided based on the dominant peak in the spin structure factor. The
color convention is as in Fig.~\ref{fig2}.
(b) DMRG results showing the dominance 
of the CX state $(\pi,0)$ 
 with increasing $L$ in the spin structure factor $S(q_x,0)$ 
using 2$\times$$L$ clusters at $U/W=0.83$,  
$J_{\rm H}/U=0.25$, and $n=2$.
(c) Same as (b) but for the block-AFM state ($\pi$/2,0),
at $U/W=0.83$, $J_{\rm H}/U=0.25$, and $n=2.5$. 
The hoppings used in all panels
are those of the pnictides (Ref.~\onlinecite{graser}).
}
\vskip -0.4cm
\label{fig4}
\end{center}
\end{figure}

\subsection{Lanczos and pnictides hoppings}

By the procedure described above, 
the results in Fig.~\ref{fig4}(a) were obtained using the Lanczos method
on a 2$\times$4 lattice, employing the original hopping amplitudes 
of the two-orbital model for the pnictides. At ``half-filling''
$n=2.0$, which is  the analog of $n=6.0$ for five orbitals, 
the CX state, one of the states observed in neutron scattering 
for ladders,\cite{caron2} dominates. 
Increasing the electronic
density $n$, the block-AFM state (the other state reported experimentally)~\cite{caron} 
is also stabilized, while decreasing $n$ the CY state 
(this state has not been reported experimentally yet) becomes the ground state.
While a quantitative agreement with the five-orbital model 
results should not be expected,
it is still reassuring that some of the main tendencies are similar when using two
and five orbitals. These common aspects are: the
block-AFM state is stable at a robust $U/W$, the CX state 
dominates a large fraction of the phase
diagram, and other states 
appear as competing alternatives.
The reduction from
five to two orbitals preserves the essence of the problem, 
at least qualitatively and at the electronic densities of the parent compounds. 
Note also that the $U/W$ needed to stabilize
the block-AFM phase is approximately 1, namely larger than in the
five-orbitals HF analysis. As explained before, 
the quantum fluctuations considered in the Lanczos calculation could
move the critical $U/W$ to larger values than in the mean-field
approximation highlighting the need to include correlation effects. 
However, since other factors such as finite size
effects in Lanczos could influence on the actual values of the critical couplings,
this conclusion should be considered only as qualitative. Moreover,
note that the comparison between two- and five-orbitals is also only
qualitative since, for instance, working at fixed $J_{\rm H}/U$
the phase diagram of Fig.~\ref{fig2} at $n = 6.0$
contains the block-AFM phase at a large enough $U/W$, while the phase diagram of
Fig.~\ref{fig4}~(a) at $n = 2.0$  only has one magnetically ordered state, the CX state. 
Another discrepancy is the dominance of the CY state for $n$ less than 2.0, feature
that is not observed in a five-orbitals context. Thus, the results of both
models are certainly not in one to one correspondence.

\subsection{DMRG and pnictides hoppings}

This same two-orbital Hubbard model was also studied 
using DMRG techniques.\cite{dmrg}
Open boundary conditions were employed in both directions, 
and typically $m=450$ states and
20 sweeps were used. The energy difference between the last two DMRG sweeps 
was $10^{-5}$. The results are in
Figs.~\ref{fig4} (b,c). In panel (b), the spin structure 
factor is shown varying $L$ using a 2$\times$$L$ cluster with $L$ = 4, 8, 
12, and 16,
at $n=2$, and fixed $J_{\rm H}/U$ and $U/W$. 
In this case, the CX state dominates,\cite{comment} 
as found with the Lanczos
technique. In panel (c), the spin 
structure factor is shown at $n=2.5$, $J_{\rm H}/U=0.25$,
intermediate coupling $U/W=0.83$, and three cluster
sizes showing that in this regime the block-AFM state dominates, as
in the Lanczos calculations. Overall, the DMRG and Lanczos results are in
qualitative agreement, showing that the CX-state 
and block-AFM state reported experimentally are stable in portions of 
the phase diagram of the two-orbital Hubbard model with the FeAs hoppings, 
while other states such as CY and FM (not shown) are close in energy.

\begin{table}
\begin{ruledtabular}
\begin{tabular}{|c|c|c|}
Matrix & 123 compound & DMRG hoppings\\ \hline
$t^x$ & 
$\begin{pmatrix}
0.14769 & -0.05309\\
0.05309 & 0.27328
\end{pmatrix}$ & 
$\begin{pmatrix}
0.14769 & 0\\
0 & 0.27328
\end{pmatrix}$ \\ \hline
$t^y$ & 
$\begin{pmatrix}
0.28805 & 0.01152\\
0.01152 & 0.00581
\end{pmatrix}$ & 
$\begin{pmatrix}
0.28805 & 0.01152\\
0.01152 & 0.00581
\end{pmatrix}$ \\ \hline
$t^{x+y}$ & 
$\begin{pmatrix}
-0.21166 & -0.08014\\
-0.08430 & -0.18230
\end{pmatrix}$ & 
$\begin{pmatrix}
-0.21166 & -0.08430\\
-0.08430 & -0.18230
\end{pmatrix}$ \\ \hline
$t^{x-y}$ & 
$\begin{pmatrix}
-0.21166 & 0.08014\\
0.08430 & -0.18230
\end{pmatrix}$ & 
$\begin{pmatrix}
-0.21166 & 0.08430\\
0.08430 & -0.18230
\end{pmatrix}$ \\
\end{tabular}
\end{ruledtabular}
\vskip 0.2cm
\caption{\label{tablehops} Set of hopping matrices 
for the two-orbital model corresponding 
to the 123-ladder compound (in eV units), obtained
from the selenides set of hoppings 
for these materials using the five-orbital model. 
The matrices are written in the orbital basis $
\{d_{xz},\,d_{yz}\}$ along the $x,\,y,\,x+y,$ and $x-y$ directions. 
The long direction of the ladder is oriented along the $x$ axis (for the five-orbitals
results, the long direction is the $y$ axis, thus a rotation was carried out).
{\it ``123 compound'' column}: hopping parameters 
obtained from tight-binding fits to the band calculations 
for BaFe$_2$Se$_3$ presented in the section of
the five-orbital Hubbard model. As explained in the five-orbitals section
the fully Hermitian matrix is obtained by constructing 4$\times$4 matrices
using the transpose of the 2$\times$2 matrices here provided.
{\it ``DMRG hoppings'' column}: 
resulting hopping set for a two-orbital model with one iron atom 
per ``unit cell'' used in the actual DMRG calculations for the 123-ladder compound. 
In this case all of the hopping matrices are Hermitian already for the 2$\times$2 cases.}
\end{table}

\subsection{Selenides hoppings for the two-orbitals model}

As explained before, the DMRG results shown above in Fig.~\ref{fig4} 
for the two-orbital Hubbard model were obtained by using 
the hopping parameters originally developed for pnictide compounds.
As also addressed for the case of the five-orbital Hubbard model, 
this is a crude approximation for BaFe$_2$Se$_3$ because the hopping parameters 
are material-dependent making it imperative a study of 
the two-orbital Hubbard model with a set of hoppings 
corresponding to the true 123-ladder selenide 
compound. This more realistic set of parameters was obtained by 
fitting tight-binding models to the band structure 
first-principles results discussed
in Sec.~II in the context of the five-orbital Hubbard model. 
Letting $t^{\alpha}_{\gamma,\gamma'}$ be the hopping matrix 
defined in the orbital space $\gamma=\{d_{xz},\,d_{yz}\}$ along the $\alpha=x,\,y,\,x+y,$ and $x-y$ 
directions between nearest and next-nearest iron atoms, 
the set of hoppings for two orbitals can be obtained, and they are
shown in the second column of Table~\ref{tablehops} (``123 compound'' column). 
Note that the five-orbitals results had the long direction along the $y$ axis,
but here a rotation was carried out and the long direction is along the $x$ axis.

For the two-leg ladder 
used in the DMRG calculations, the presence of the Se 
atoms with a staggered location above and
below the FeSe ladders in the 123 material imply that the ``unit cell'' must
be larger than in the absence of that lattice distortion.\cite{staggered-comment} 
The other small lattice distortion along the ladder legs
incorporated in the five-orbitals results also demand such an enlarged unit cell. 
However, this doubling of the unit cell increases 
the complexity of the DMRG calculation because operators for the additional 
iron atoms belonging 
to the enlarged unit-cell must be kept. 
Then, in order to simplify the DMRG computation, 
approximations (discussed in the next paragraph) 
will be introduced to reduce the problem
to a two-orbital Hubbard model where all Fe-Fe bonds along the leg direction
are equivalent.
The final set of hoppings used in our DMRG calculation 
are shown in the ``DMRG hoppings'' 
column shown in Table~\ref{tablehops}. They were obtained by averaging
hoppings along the same direction but for different bonds. 


\begin{figure}
\centering
\includegraphics[width=8.5cm]{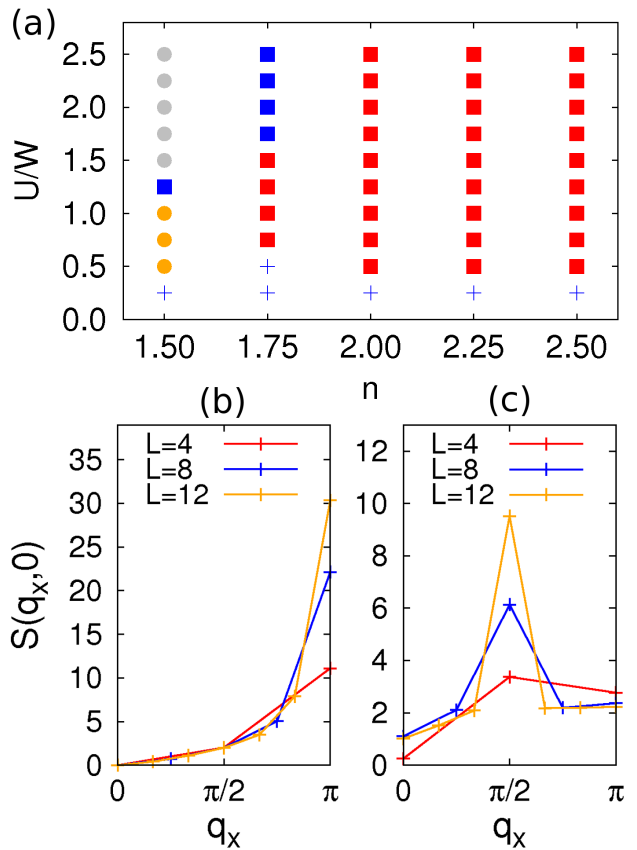}
\caption{(Color online) (a) Phase diagram of the two-orbital Hubbard model studied 
with the DMRG method employing the hopping 
amplitudes presented in Table~\ref{tablehops} (right column). 
The cluster used is of size $2\times 8$ with open boundary conditions, 
and $J_{\rm H}/U=0.25$. Shown are results varying $U/W$ 
and the number of total electrons $N$ as 24, 28, 32 ($n=2$), 36, and 40. 
The assignments of the many phases are decided based on the wavevector of the 
dominant peak in the spin structure factor. The ``PM'' (paramagnetic) phase was
assigned based on the similarity with the $U=0$ results, and for this reason 
the frontier
between the PM and  magnetic phases may  not be precise. 
The color convention is as in Fig.~4, and 
300 states and 15 sweeps were used. (b) DMRG results showing the dominance 
of the CX state $(\pi,0)$ with increasing $L$ in the spin structure factor 
$S(q_x, 0)$ using $2\times L$ clusters at $U/W=1.25$, 
$J_{\rm H}/U=0.25$, and $n=2$. In these runs, 450 states and 15 sweeps were used. 
(c) Same as (b) 
but for the block-AFM state with wavector $(\pi/2,0)$, at $U/W=1.25$, 
$J_{\rm H}/U=0.25$, and $n=1.5$, 
using $2\times 4$, $2\times 8$, and $2\times 12$ clusters.}
\label{fig.5.dmrg}
\end{figure}

\subsection{DMRG and selenides hoppings}

The results obtained with the DMRG technique applied to the two-orbital
Hubbard model using the hoppings for selenides described in the previous
subsection are shown in Fig.~\ref{fig.5.dmrg}. Technical details are
the same as in the case of the pnictides hoppings. Panel (a) contains the phase
diagram from the wavevectors that dominate in the spin structure factor.
Panels (b,c) display the behavior of the spin structure factor at particular
couplings, for the two most important states.
Similarly as in the case of the hoppings for pnictides, at $n=2$ there
is a dominance of the CX state. In fact, the region of stability of this
CX state is larger with the selenides hoppings than with the pnictides
hoppings. This may lead to the conclusion that for new ladder compounds
synthesized in the future the CX state should be more likely to appear than
the block-AFM state. Or it could be 
that our analysis does not include a lattice distortion that favors the
block-AFM state over the CX-state. 

With regards to the block-AFM 
state, in Fig.~\ref{fig.5.dmrg}~(a) it is shown that
this phase indeed exists in the regime of hole doping, contrary to the case
of the hoppings for pnictides where the block-AFM 
state was found for electron doping,
in a region of approximately the same size in the phase diagram. 
Not finding this state at precisely $n=2$ is not a problem since the actual population
of the $d_{xz}$ and $d_{yz}$ orbitals is not precisely 2 in the real materials. What
is perhaps more surprising is the dominance of the CX-state 
over the block-AFM state, contrary to the results of the five-orbitals model
where the latter was fairly stable. This result highlights the shortcomings
of the two-orbitals model.

In the phase diagram of Fig.~\ref{fig.5.dmrg} other phases not observed
experimentally are found such as
the CY-state and the FM-state, again similar to the case of the  hoppings
for pnictides or for the five-orbitals models. While the precise location of
the phases varies from model to model and depends on the hoppings, the systematic
presence of the block-AFM and CX states is clear in our results.

However, as in the case of the pnictides hoppings, it is important to remark
that the agreement between five- and two-orbitals calculations is 
qualitative at best. For instance, consider the case of $n = 2.0$ and
$J_{\rm H}/U = 0.25$ in 
Fig.~\ref{fig.5.dmrg}~(a). Here, as in the case of the pnictides hoppings
Fig.~\ref{fig4}~(a), the CX state dominates in the range of $U/W$ explored.
However, for the case of five-orbitals at $n = 6.0$ and $J_{\rm H}/U = 0.25$ 
in Fig.~\ref{fig2}~(a), it is the block-AFM state that dominates. Thus, only
by including variations in the electronic density $n$ is that similarities
between the two cases do emerge.

\subsection{Spin fermion model results} 

Complementing these studies, the two-orbital spin-fermion model 
for the pnictides~\cite{spin-fermion1,spin-fermion2} 
was also analyzed using a 2$\times$16 cluster
and Monte Carlo (MC) techniques, employing approximately 50,000 MC steps (not shown). 
At $n \sim 2$ and $J_{\rm H} \geq 0.2$~eV, the CX state was found, while 
at $n=2.5$ and a similar range of $J_{\rm H}$ the block-AFM state 
was found, in agreement
with the Lanczos and DMRG studies. 
Results will be presented in future publications.

\section{Conclusions} 

Using five- and two-orbital models for the two-leg ladder compounds
BaFe$_2$Se$_3$ and KFe$_2$Se$_3$, the phase diagram of these models were studied
using several many-body techniques. The richness
of the reported phase diagrams
demonstrates that Fe-based superconductors are more complex 
than early investigations based on Fermi Surface nesting ideas 
anticipated.\cite{recent-review}
More specifically, in this study it has been argued 
that the experimentally observed CX 
and 2$\times$2 block-AFM states shown in Fig.~\ref{fig1} 
are indeed the ground state of purely electronic Hubbard models in
robust regions of parameter space when varying $U/W$, 
$J_{\rm H}$, and the electronic density $n$, at least 
within the HF approximation. 
Our effort suggests that to understand the stability 
of the 2$\times$2 block states, theoretical
studies of electronic models using the geometry of  two-leg ladders 
(much simpler than a full two-dimensional layer)
may be sufficient, although for a quantitative 
description quantum fluctuations and the effect of lattice
distortions may be needed.
Our study also predicts that several other 
magnetic phases could become stable in the vicinity of the CX and block-AFM states
in the phase diagram. The other candidates 
are in Fig.~\ref{fig1} and the list includes
the G-AFM, CY, and FM states, and to a lesser 
extent the Flux and  T states. The experimental search for
these states via chemical substitution or pressure 
would be important to improve the
interplay between theory and experiments for the Fe-based superconductors. 
Since these magnetic arrangements 
are close in energy, glassy behavior
caused by a multiplicity of energy minima is also possible.\cite{athena} 
Finally, by comparing results using two sets 
of hopping amplitudes (one realistic for
the ladder selenides and obtained via first-principles 
calculations, and another borrowed
from pnictides investigations), several 
similarities were unveiled particularly at intermediate
and large Hubbard couplings.

\section{Acknowledgments} 

The authors thank A. S. Sefat and B. Saparov for useful conversations.
This work was supported by 
the U.S. DOE, Office of Basic Energy Sciences,
Materials Sciences and Engineering Division (Q.L, S.L., G.D.S., E.D.), by the National
Science Foundation under Grant No. DMR-1104386 (A.N., Rinc\'on, A.M.), by
CONICET, Argentina (Riera), and by the Center for
Nanophase Materials Sciences, sponsored by the Scientific User Facilities 
Division, BES, U.S. DOE (G.A.).
L.W. and W. K. acknowledge support by the U. S. Department 
of Energy, Office of Basic Energy Sciences
DE-AC02-98CH10886, DE-FG02-05ER46236, and DOE-CMCSN.


\begin{thebibliography}{10}
\bibitem{comment} Whether a true long-range
order or a slow power-law decay in the correlation functions dominates is difficult
to establish with the limited $L$ data set available.
\bibitem{staggered-comment}
For the FeAs layers this staggering 
is also present, but in the ``pnictides'' hoppings used in this publication the effect was ignored,
i.e. all the As atoms were assumed to be at the same height with respect to the Fe layer.
\end{thebibliography}
\end{document}